\documentclass{JHEP3}
\usepackage{amsmath}
\usepackage{cite}
\usepackage{epsfig}
\usepackage{xspace}
\usepackage{graphics}
\usepackage[latin1]{inputenc}

\def\hide#1{}

\newcommand{\as}{\ensuremath{\alpha_{\mathrm{s}}}}

\newcommand{\gtaet}{\raisebox{-0.8mm}%
{\hspace{1mm}$\stackrel{>}{\sim}$\hspace{1mm}}}

\newcommand{\lam}{\ensuremath{\Lambda_{\mathrm{QCD}}}}

\def\mrm#1{\mathrm{#1}}

\def\sub#1{\ensuremath{_{\mrm{#1}}}}

\def\f2d3{\ensuremath{F_2^{\mrm{D}3}}}

\def\done#1{}

\providecommand{\eqref}[1]{eq.~(\ref{#1})\xspace}
\renewcommand{\eqref}[1]{eq.~(\ref{#1})\xspace}

\newcounter{aenumct}

\newcounter{ienumct}

{\end{list}}

\newcounter{enumct}

\usepackage{ucs}
\PrerenderUnicode{������}

\newcommand{\si}[0]{\ensuremath{\sigma}}
\newcommand{\te}[0]{\ensuremath{\theta}}
\newcommand{\ph}[0]{\ensuremath{\phi}}
\newcommand{\abs}[1]{\ensuremath{\left| #1 \right|}}
\newcommand{\Ps}[0]{\ensuremath{\Psi}}
\newcommand{\ga}[0]{\ensuremath{\gamma}}
\newcommand{\la}[0]{\ensuremath{\lambda}}

\newcommand{\de}[0]{\ensuremath{\delta}}
\newcommand{\al}[0]{\ensuremath{\alpha}}
\newcommand{\ep}[0]{\ensuremath{\varepsilon}}
\newcommand{\bl}[1]{\mathbf{#1}}
\newcommand{\om}[0]{\ensuremath{\omega}}

\newcommand{\rh}[0]{\ensuremath{\rho}}

\newcommand{\rmax}[0]{\ensuremath{r_{\max}}}

\usepackage{axodraw}
\usepackage{epsfig}
\usepackage{amssymb}
\usepackage{mathrsfs}
\usepackage{mathcomp}
\usepackage{cite}
\usepackage[latin1]{inputenc}

\def\pmb#1{{\mbox{\boldmath$#1$}}}

\preprint{MCnet/08/05\\
  LU-TP 08-14
}

\title{Elastic and quasi-elastic $pp$ and $\gamma^\star p$ scattering
  in the Dipole Model\footnote{Work supported in part by the Marie
    Curie RTN ``MCnet'' (contract number MRTN-CT-2006-035606), by the
    Swedish Foundation for International Cooperation and Higher
    Education -- STINT (contract number 2006/080) and by the Deutsche
    Forschungsgemeinschaft.}}

\author{Christoffer Flensburg$^\dag$, G\"osta Gustafson$^{\dag\S}$ and
  Leif L\"onnblad$^\dagger$\\
  $^\dag$Dept.~of Theoretical Physics,
  S\"olvegatan 14A, S-223 62  Lund, Sweden\\
  $^\S$II. Institut f\"ur Theoretische Physik, Universit\"at Hamburg,
  Luruper Chaussee 149, 22761 Hamburg, Germany\\
  E-mail: \email{christoffer.flensburg@thep.lu.se},
  \email{gosta.gustafson@thep.lu.se} and \email{leif.lonnblad@thep.lu.se}}

\abstract{We have in earlier papers presented an extension of
  Mueller's dipole cascade model, which includes sub-leading effects
  from energy conservation and running coupling as well as colour
  suppressed saturation effects from pomeron loops via a ``dipole swing''. The
  model was applied to describe the total and diffractive cross
  sections in $pp$ and $\gamma^*p$ collisions, and also the elastic cross
  section in $pp$ scattering.

  In this paper we extend the model to describe the corresponding
  quasi-elastic cross sections in $\gamma^*p$, namely the exclusive
  production of vector mesons and deeply virtual compton scattering.
  Also for these reactions we find a good agrement with measured cross
  sections. In addition we obtain a reasonable description of the
  $t$-dependence of the elastic $pp$ and quasi-elastic $\gamma^\star
  p$ cross sections.}

\begin{document}

\sloppy

\section{Introduction}
\label{sec:intro}

We have in a series of papers
\cite{Avsar:2005iz,Avsar:2006jy,Avsar:2007xg} presented an extention
of Mueller's dipole cascade model
\cite{Mueller:1993rr,Mueller:1994jq,Mueller:1994gb} implemented in a
Monte Carlo program, which includes sub-leading effects from energy
conservation and running coupling, as well as colour suppressed
effects from pomeron loops via a \textit{dipole swing} mechanism. It also 
includes a consistent treatment of non-perturbative 
confinement effects, which suppress dipoles with
large transverse extension.

The advantage of a cascade model formulated in transverse coordinate 
space is the possibility to include effects of multiple collisions and 
saturation in a straight forward way. While analytic results have mainly been 
presented for the asymptotic behavior of total and diffractive cross 
sections, Monte Carlo simulations facilitate
studies of non-leading effects and more quantitative results. A simulation of 
Mueller's initial model was presented by Salam in ref.~\cite{Salam:1996nb}.
Although giving finite results for the total cross section, this leading 
log evolution suffers from divergences for small 
dipoles, which caused numerical problems with very large gluon 
multiplicities and prevented simulations at higher energies. 
One important result from this analysis was the very large 
fluctuations in the evolution \cite{Mueller:1996te}. As the 
ratio between the elastic and the total cross sections is determined by
the fluctuations in the scattering process, this implies that less 
fluctuations is needed in the impact parameter dependence, to reproduce the
experimental data. As a result we found in ref.~\cite{Avsar:2007xg} that 
including the fluctuations in the evolution implies that the impact 
parameter profile is not as ``black and white'' as in analyses where only
fluctuations in the impact parameter are taken into account. 

In the model described in
refs.~\cite{Avsar:2005iz,Avsar:2006jy,Avsar:2007xg} we include a
number of sub-leading effects, with the aim that we in the end will be
able to describe not only the total and diffractive cross sections,
but also to generate fully exclusive final states.  The main ingredient
in our model is energy conservation, which is included by assigning a
transverse momentum to each emitted gluon given by the maximum inverse
size of the neighboring dipoles. As a result this also implies that
the singularities for small dipoles are avoided. Other features are
saturation effects in the evolution through a dipole swing mechanism,
and a consistent treatment of confinement and running coupling effects
in both dipole emissions and dipole--dipole interactions.

Taken together with a very simple model for the initial proton
wavefunction, these features allow us to obtain a Lorentz-frame
independent description of total cross sections, both for $pp$ and
DIS, using basically only two free parameters, a confinement scale \rmax\ and
$\lam$\cite{Avsar:2007xg}.  The model gives a good
description of measurements of the total and diffractive cross
sections in $pp$ and $\gamma^*p$ collisions, and also for the elastic
cross section in $pp$.
In this paper we will continue our investigations with an analysis of 
exclusive production of vector mesons and real photons in $\gamma^*p$. 
The aim is to further test our model, and in
particular to study the effect of the fluctuations in the cascade. We
also extend the analyses to include the $t$-dependence of the
(quasi-)elastic cross sections, including also elastic $pp$ scattering,
which in particular gives information about the
properties of the incoming proton state.

In the eikonal approximation the quasi-elastic $\gamma^* p$ collisions
contain three elements: the virtual photon--dipole vertex, the
dipole--proton scattering amplitude, and the vertex for the transition
between the dipole and the final vector meson or real photon. Here the
first component can be calculated perturbatively, although a hadronic
component must be included at lower $Q^2$-values. In an extensive
study Forshaw \emph{et al.}\ \cite{Forshaw:2003ki, Forshaw:2006np} have
analyzed the results obtained from a set of models for the
dipole--proton scattering and for the vector meson wavefunctions, and
compared them with experimental data. In this paper we want to carry
out a similar analysis, but now use our dipole cascade model for the
dipole--proton scattering.  We are here particularly interested in
effects of fluctuations in the cascade evolution, which are not
included in the analyses by Forshaw \emph{et al.}  We also want to
use this study to put constraints on the state of the incoming proton.

We begin in section \ref{sec:formalism} with discussing the eikonal
formalism for exclusive vector meson production, whereafter we
describe our model for dipole evolution and dipole--dipole scattering
in section \ref{sec:dip} and the models we use for the
proton, photon and vector meson wavefunctions in section
\ref{sec:wfs}. In section \ref{sec:tuning} we retune the parameters of
our model to data on total and elastic $pp$ cross sections and the
total $\gamma^\star p$ cross section before we present our results on
quasi-elastic $\gamma^\star p$ cross sections on section
\ref{sec:results-quasi-elast}. Finally we present our conclusions in
section \ref{sec:conc}

\section{Formalism}
\label{sec:formalism}

\subsection{The dipole cascade model and the eikonal approximation}

As discussed in the introduction, our model for $pp$ collisions and DIS
is an extension of Mueller's dipole cascade model
\cite{Mueller:1993rr,Mueller:1994jq,Mueller:1994gb}. In this formalism the
probability per unit rapidity $Y$ that a dipole $(\pmb{x},\pmb{y})$
emits a gluon at transverse position $\pmb{z}$ is given by
\begin{eqnarray}
\frac{d\mathcal{P}}{dY}=\frac{\bar{\alpha}}{2\pi}d^2\pmb{z}
\frac{(\pmb{x}-\pmb{y})^2}{(\pmb{x}-\pmb{z})^2 (\pmb{z}-\pmb{y})^2},
\,\,\,\,\,\,\, \mathrm{with}\,\,\, \bar{\alpha} = \frac{3\alpha_s}{\pi}.
\label{eq:dipkernel1}
\end{eqnarray}
The evolution of this cascade agrees with the leading order BFKL
evolution. As a consequence, the total number of dipoles grows
exponentially. This also implies a strong growth for the total cross
section which, however, is tamed by taking multiple dipole
interactions into account. The scattering probability between two
elementary colour dipoles with coordinates $(\pmb{x}_i,\pmb{y}_i)$ and
$(\pmb{x}_j,\pmb{y}_j)$ respectively, is given by
\begin{equation}
  f_{ij} = f(\pmb{x}_i,\pmb{y}_i|\pmb{x}_j,\pmb{y}_j) =
  \frac{\as^2}{8}\biggl[\log\biggl(\frac{(\pmb{x}_i-\pmb{y}_j)^2
    (\pmb{y}_i-\pmb{x}_j)^2}
  {(\pmb{x}_i-\pmb{x}_j)^2(\pmb{y}_i-\pmb{y}_j)^2}\biggr)\biggr]^2.
\label{eq:dipamp}
\end{equation}
Summing over all $i$ and $j$ this can give a large number, but the
unitarized interaction probability determined by
\begin{equation}
T(\mathbf{b}) = 1-e^{-\sum f_{ij}} \equiv 1-e^{-F}
\end{equation}
will never exceed 1.

We note that the splitting probability in \eqref{eq:dipkernel1} is
singular for small dipole sizes $\pmb{x-z}$ or $\pmb{z-y}$, but
these small dipoles have a small probability to interact with the
target, and the eikonal $F=\sum f_{ij}$ is finite.

In the model developed in refs.~\cite{Avsar:2005iz,Avsar:2006jy,Avsar:2007xg},
we extended Mueller's cascade model to include sub-leading effects from energy
conservation and a running coupling,
saturation effects not only in the dipole--dipole subcollisions
but also within the individual cascades, and effects of
confinement. These features are further discussed in sec.~\ref{sec:dip}.

The model is supplemented by a non-perturbative model for an
initial proton in terms of three dipoles.
In the eikonal approximation the total and the diffractive 
(including the elastic) cross sections are then given by
\begin{eqnarray}
\sigma\sub{tot}= 2\int\! d^2 b \,\langle (1-e^{-F})\rangle,\,\,\,\,
\sigma\sub{diff}= \int \!d^2 b\,\langle (1-e^{-F})^2\rangle.
	\label{eq:eikonalCS}
\end{eqnarray}
The diffractive cross section can be separated in elastic scattering
and diffractive excitation:
\begin{eqnarray}
\sigma\sub{el}= \int\! d^2 b\,(\langle 1-e^{-F}\rangle)^2,\,\,\,\,
\sigma\sub{diff\,exc}= \int \!d^2 b\,\left\{\langle (1-e^{-F})^2\rangle -
(\langle 1-e^{-F}\rangle)^2 \right\}.
	\label{eq:elasticCS}
\end{eqnarray}
Thus the separation between elastic and inelastic diffraction is 
determined by the fluctuations in the scattering amplitude.
The average in eqs.~(\ref{eq:eikonalCS}) and 
(\ref{eq:elasticCS}) is taken over the different incoming dipole
configurations and different cascade evolutions,
which thus give two separate sources for fluctuations.
The differential cross section $d \sigma\sub{el}/d t$ is obtained
from the Fourier transform of the scattering amplitude
\begin{equation}
\frac{d \sigma_\lambda}{d t} = \frac{1}{4 \pi} 
\left|\int d^2 \mathbf{b}\, e^{i \mathbf{q}\mathbf{b}} \langle 1-e^{-F}\rangle 
\right|^2,
\,\,\,\,\,\,\,\mathrm{with}\,\,\, t=-\mathbf{q}^2.
\label{eq:dsigma}
\end{equation}

\subsection{DVCS and exclusive vector meson production in $\gamma^*p$
  collisions} 

We want to study the 
exclusive processes
\begin{equation}
\gamma^* \, p \rightarrow V\, p,\,\,\,\, V=\gamma,\rho,\psi, \ldots
\end{equation}
In the dipole model the virtual photon is split into a $q\bar{q}$ pair
long before the collision. This dipole
scatters elastically against the proton, and after the scattering the pair
joins again forming a real photon or a vector meson.
The formulation in the transverse coordinate plane makes it easier to 
study these pseudo-elastic reactions, and in the eikonal approximation 
the scattering amplitude is expressed in terms of three components:
\begin{equation}
  \Im A_\lambda(s,\mathbf{b}) = s \sum_{f,h,\bar{h}} \int \int dz \,
  d^2\mathbf{r}\,
  \Psi_{f h \bar{h}}^{* V \lambda}(\mathbf{r},z)\,
  \Psi_{f h \bar{h}}^{\gamma \lambda}(\mathbf{r},z,Q^2)\,
  \hat{\sigma}_{dp}(s,\mathbf{r},\mathbf{b},z).
\label{eq:amplitude}
\end{equation}
Here $\mathbf{r}$ is the transverse size of the dipole, $z$ and $1-z$
the fractions of the photon or vector meson carried by the quark and
antiquark respectively, and $h$ and $\bar{h}$ their
helicities. $\lambda$ denotes the photon or vector meson helicity,
$\hat{\sigma}_{dp}$ is the dipole--proton scattering probability with
$\mathbf{b}$ the impact parameter, and $s$ the total energy squared.

Neglecting the small contribution from the real part of the amplitude, 
the total cross section is given by
\begin{equation}
\sigma_{\lambda}(\gamma^* p \rightarrow V p) = \frac{1}{4 s^2} 
\int d^2 \mathbf{b}\, |A_\lambda(s,\mathbf{b})|^2.
\label{eq:elastic}
\end{equation}
In analogy with \eqref{eq:dsigma} the differential cross section is 
obtained from the Fourier transform:
\begin{equation}
\frac{d \sigma_\lambda}{d t} = \frac{1}{16 \pi s^2} 
\left|\int d^2 \mathbf{b}\, e^{i \mathbf{q}\mathbf{b}}  
A_\lambda(s,\mathbf{b})\right|^2,
\,\,\,\,\,\,\,\mathrm{with}\,\,\, t=-\mathbf{q}^2.
\label{eq:diffelastic}
\end{equation}

\section{The improved dipole cascade}
\label{sec:dip}

As discussed in the introduction, the model developed in
refs.~\cite{Avsar:2005iz,Avsar:2006jy,Avsar:2007xg}.
is an extension of Mueller's dipole
cascade model, which includes sub-leading effects from energy
conservation and a running coupling,
saturation effects not only in the dipole--dipole subcollisions
but also within the individual cascades, and effects of confinement. 
As mentioned above, an essential point is here that we include the effect of
fluctuations in the dipole cascades in the calculation of the elastic
or quasi-elastic cross sections.

\subsection{Non-leading perturbative effects}
\subsubsection{Energy-momentum conservation}

It is known that the large NLO corrections to the BFKL evolution are
reduced significantly if proper energy conservation is included in the
leading order.  In our model a small transverse extension is
interpreted as a large transverse momentum. This interpretation is
supported by the resulting analogies between the dipole chains in
coordinate space and the chains in the LDC model, which is formulated
in momentum space and interpolates smoothly between DGLAP and BFKL
evolution.  Taking energy-momentum conservation into account is most
easily done in a Monte Carlo (MC) simulation.  Conserving both
light-cone components, $p_+$ and $p_-$, implies that we also satisfy
the so called consistency constraint\cite{Kwiecinski:1996td}. As small
dipoles in our formalism correspond to large transverse momenta,
energy conservation also gives a dynamical cutoff for the otherwise
diverging number of small dipoles, and thus makes the MC simulation
much more efficient.

\subsubsection{Running coupling}

In our simulations we also include non-leading effects from the
running of \as, both in the dipole splitting and in the dipole--dipole
scattering probability. In the dipole emissions the scale in the
coupling is given by $\min(r, r_1, r_2)$, where $r$ is the size of the
mother dipole which splits into $r_1$ and $r_2$. This is the most natural
choice and is also consistent with recent NLO calculations
\cite{Balitsky:2006wa,Kovchegov:2006wf,Balitsky:2007}.
For the dipole--dipole scattering the situation is somewhat more
complicated with basically six different dipole sizes involved. We
have chosen to use the scale $\min(|\pmb{x}_i -
\pmb{y}_i|,|\pmb{x}_j-\pmb{y}_j|, |\pmb{x}_i-\pmb{y}_j|,
|\pmb{y}_i-\pmb{x}_j|)$.
In order to avoid divergencies the coupling is in all cases frozen so that
$\as(r)\to\as(\min(r,r_{\max}))$, where $r_{\max}$ is the confining scale
discussed in section \ref{sec:confinement} below.
 
\subsection{Saturation within the cascades} 

Mueller's cascade includes saturation effects from multiple collisions
in the Lorentz frame chosen for the calculation, but not saturation
effects from gluon interaction within the individual cascades. The
result is therefore dependent on the chosen Lorentz frame. In
ref.~\cite{Avsar:2006jy} we improved our model by allowing (colour
suppressed) recouplings of the dipole chain during the evolution, a
``dipole swing''.  The swing is a process in which two dipoles
$(\pmb{x}_i,\pmb{y}_i)$ and $(\pmb{x}_j,\pmb{y}_j)$ are replaced by
two new dipoles $(\pmb{x}_i,\pmb{y}_j)$ and
$(\pmb{x}_j,\pmb{y}_i)$. The process can be interpreted in two
ways. There is a probability $1/N_c^2$ that the two dipoles may have
the same colour, and the quark at $\pmb{x}_i$ and the antiquark at
$\pmb{y}_j$ form a colour singlet. In this case the best approximation
of the quadrupole field ought to be obtained by the closest
charge-anticharge combinations.  Here the swing is therefore naturally
suppressed by $1/N_c^2$, and it should be more likely to replace two
given dipoles with two smaller ones. Secondly, we may see it as the
result of a gluon exchange between the dipoles, which results in a
change in the colour flow. In this case the swing would be
proportional to $\as^2$, which again is formally suppressed by
$N_c^2$, compared to the splitting process in \eqref{eq:dipkernel1},
which is proportional to $\bar{\alpha}=N_c \as/\pi$.

In the MC implementation each dipole is randomly given one of $N_c^2$
possible colour indices. Only dipoles with the same colour can swing, and
the weight for a swing
$(\pmb{x}_1,\pmb{y}_1),(\pmb{x}_2,\pmb{y}_2)\rightarrow
(\pmb{x}_1,\pmb{y}_2), (\pmb{x}_2,\pmb{y}_1)$ is determined by a factor
proportional to
\begin{equation}
\frac{(\pmb{x}_1-\pmb{y}_1)^2 (\pmb{x}_2-\pmb{y}_2)^2}
{(\pmb{x}_1-\pmb{y}_2)^2 (\pmb{x}_2-\pmb{y}_1)^2}.
\end{equation}
This implies that the swing favors the formation of smaller dipoles.
The number of dipoles is not reduced by the swing, but the fact that
smaller dipoles have smaller cross sections gives the desired
suppression of the total cross section. Although not explicitely frame
independent the results from the MC simulations are very nearly
independent of the Lorentz frame used for the calculations.

\subsection{Confinement effects}
\label{sec:confinement}

Mueller's dipole model is a purely perturbative process. It should therefore 
be applied to small dipoles, \emph{e.g.} to heavy quarkonium states.
When applying the dipole formalism to collisions with protons it is
necessary to take confinement into account, in order to prevent
the formation of very large dipoles. Confinement effects must also
suppress long range interactions between colliding dipoles. 
In ref.~\cite{Avsar:2007xg} a
consistent treatment of confinement was presented by replacing the Coulomb
potentials in eqs.~(\ref{eq:dipamp}) and (\ref{eq:coulombprop}) by
screened potentials, with a screening length $r_{\max}$.

 Obviously the dipoles produced in
the splitting process in \eqref{eq:dipkernel1} cannot become too
large, and it is natural to introduce a scale $r_{\max}$, so that
larger dipoles are suppressed. In a similar way confinement must
suppress long range interactions between colliding dipoles.

The formula for $f_{ij}$ in \eqref{eq:dipamp} is
just the two dimensional Coulomb potential, and can be written as 
\begin{eqnarray}
  f(\pmb{x}_i,\pmb{y}_i|\pmb{x}_j,\pmb{y}_j) =
  \frac{g^4}{8} [ \Delta(\pmb{x}_i - \pmb{x}_j)
  - \Delta(\pmb{x}_i - \pmb{y}_j) - \Delta(\pmb{y}_i - \pmb{x}_j) +
  \Delta(\pmb{y}_i - \pmb{y}_j) ]^2\label{eq:ddgreen}
\end{eqnarray}
where $\Delta(\pmb{r})$ is the Green's function given by 
\begin{eqnarray}
  \Delta(\pmb{r}) = \int \frac{d^2\pmb{k}}{(2\pi)^2}
  \frac{e^{i\pmb{k} \cdot\pmb{r}}}{\pmb{k}^2}.
\label{eq:coulombprop}
\end{eqnarray}
To take confinement into account we replace the infinite range Coulomb
potential with a screened Yukawa potential. This implies that the
Coulomb propagator $1/\pmb{k}^2$ in \eqref{eq:coulombprop} is replaced
by $1/(\pmb{k}^2+M^2)$, where $M=1/r_{\max}$ is the confinement
scale. As a result, the four functions $\Delta$ in
\eqref{eq:ddgreen} will be replaced by
\begin{eqnarray}
  \int \frac{d^2\pmb{k}}{(2\pi)^2} \,
  \frac{e^{i\pmb{k} \cdot\pmb{r}}}{\pmb{k}^2 + 1/r_{\max}^2} 
  = \frac{1}{2\pi}K_0(r/r_{\max})
\end{eqnarray} 
with $K_0$ a modified Bessel function.
For small separations, where $r \ll r_{\max}$, the function
$K_0(r/r_{\max})$ behaves like $\ln(r_{\max}/r)$, and we then recognize
the result in \eqref{eq:dipamp}.  For large separations, $r
\gg r_{\max}$, $K_0(r/r_{\max})$ falls off exponentially $\sim
\sqrt{\frac{\pi r_{\max}}{r}}e^{-r/r_{\max}}$, as expected from
confinement.

In a similar way, the underlying Coulomb potential in the dipole
splitting function in \eqref{eq:dipkernel1} can be replaced by a
screened Yukawa potential, using again the replacement
$1/\pmb{k}^2\to1/(\pmb{k}^2+1/r^2_{\max})$. The modified splitting 
probability is then given by
\begin{eqnarray}
  \frac{d\mathcal{P}}{dY} \to \frac{\bar{\alpha}}{2\pi}d^2\pmb{z} \biggl \{
  \frac{1}{r_{\max}}\frac{\pmb{x}-\pmb{z}}{|\pmb{x}-\pmb{z}|}
  \,K_1\!\left(\frac{|\pmb{x}-\pmb{z}|}{r_{\max}}\right) - 
  \frac{1}{r_{\max}}\frac{\pmb{y}-\pmb{z}}{|\pmb{y}-\pmb{z}|}
  \,K_1\!\left(\frac{|\pmb{y}-\pmb{z}|}{r_{\max}}\right)
  \biggr \}^2. \nonumber \\
\label{eq:moddipkernel}
\end{eqnarray}
For small arguments $K_1(r/r_{\max}) \approx r_{\max}/r$, from
which we get back the result in \eqref{eq:dipkernel1}, while for large
arguments $K_1(r/r_{\max}) \sim \sqrt{\pi
r_{\max}/r}\,\cdot\,e^{-r/r_{\max}}$, and once again we obtain an
exponentially decaying field.

\section{Initial wave functions}

\label{sec:wfs}

\subsection{Proton wave function}
\label{sec:inprmo}

In ref.~\cite{Avsar:2006jy} we also introduced a simple model for the 
proton in terms of three dipoles with extensions determined by a Gaussian
distribution. The resulting model was in good agreement with total cross
sections for both DIS and $pp$ collisions.
It was shown in \cite{Praszalowicz:1997nf} that the three valence
quarks in a proton emits gluons in transverse space with the same
distribution as three dipoles, only with half the intensity. Thus, by
modeling the proton with a closed chain of three gluons we emulate the
fact that a proton at rest may contain more charges than its valence
quarks. This is analogous to the finding of Gl{\"u}ck, Reya and Vogt,
who needed a large valence-like gluon component when trying to fit
parton densities evolved from a very low scale\cite{Gluck:1998xa}.
Thus, although not a fully realistic description of the initial
non-perturbative proton state, the model appears to give a fair representation
of the multi-dipole system obtained at the low $x$-values, which are 
important for the high energy collisions.

The results turned out to be almost independent of the shape of
the three starting gluons, except for the size of the triangle. In
fact, equilateral triangles that were allowed to vary in only size and
orientation turned out to model the proton as well as more complicated
formations. With a Gaussian distribution for the size of equilateral 
triangles, motivated by the exponential dependence on $t$ for the elastic
cross section, data on total cross sections for DIS and $pp$ collisions
are well reproduced, when the width of the
Gaussian was tuned to $3.5$~GeV$^{-1} \approx 0.66$~fm.

As discussed above, the differential and elastic cross sections are determined
by the fluctuations in the scattering amplitude, and
in ref.~\cite{Avsar:2007xg} it was pointed out that a Gaussian wavefunction
as discussed above must overestimate the fluctuations of
the incoming state in its rest frame. 
The probability for the three quarks to simultaneously be located in a 
single point ought to be suppressed, and it was emphasized that the
exponential $t$-dependence of the elastic cross section, which motivated
the Gaussian shape, is only observed for $|t| < 0.15\, \mathrm{GeV}^2$,
corresponding to $b \gtaet 1$~fm.
A wavefunction of the form 
\begin{equation}
|\Psi|^2 = C\, e^{-(r-R_p)^2/w^2}
\label{eq:pwave}
\end{equation}
was also tested, and found to give essentially
identical total cross sections. The fluctuations are here suppressed by a
small value of $w$, and in ref.~\cite{Avsar:2007xg} it was observed that
reducing the fluctuations to a minimum gave good agreement with the 
integrated elastic and diffractive cross sections in $pp$ collisions. 
Lacking further constraints we
could, however, only present an upper limit for $\sigma\sub{el}$, by neglecting
the fluctuations in the wavefunction, thus including only those in the cascade
evolution. 

An essential motivation for the present analysis of quasi-elastic $\gamma^* p$ 
cross sections and of the $t$-dependence of the $pp$ elastic cross section, 
is to check whether the fluctuations in the dipole cascade model 
are also consistent with these observables, and if more constraints
can be put on the shape of the initial proton state. In this analysis we will
use the two-parameter form in \eqref{eq:pwave}, and see if this can be
adjusted to reproduce also the (quasi-)elastic cross sections.

At this point we also note that in many analyses the fluctuations in
the cascade evolution are neglected. This means that $e^{-F}$ is
replaced by $e^{-<F>}$ in eqs.~(\ref{eq:eikonalCS}) and (\ref{eq:elasticCS}).   
Including also the fluctuations in the cascade implies
that the impact parameter profile has to be more ``gray'' and less  
``black and white''. As 
an example the amplitude $\langle T(b=0)\rangle$ is a factor 2/3  
smaller in our formalism than in the analysis by Kowalski and Teaney 
\cite{Kowalski:2003hm}, for a dipole of size $2\, \mathrm{GeV}^{-1}$ and  
$x=10^{-4}-10^{-5}$. 

\subsection{Photon wavefunction}
\label{sec:phwfcn}

\subsubsection{Large $Q^2$}

For large $Q^2$ the coupling of the $\gamma^*$ to the $q\bar{q}$ pair can be 
calculated perturbatively.  The well known result to leading order is 
\begin{eqnarray}
  \Ps^{\ga 0}_{f h \bar{h}}(Q,r,z) &=& \frac{\sqrt{ \al_{EM} N_C}}{\pi} Q z (1-z) e_f K_0 (r \ep_f ) \de_{h \bar{h}}  \nonumber\\
  \Ps^{\ga +}_{f h \bar{h}}(Q,r,z) &=& \frac{\sqrt{ \al_{EM} N_C / 2}}{\pi} e_f \Big( ie^{i\te} \left(  z \de_{h+} \de_{\bar{h}-} - (1-z) \de_{h-} \de_{\bar{h}+}  \right) \ep_f K_1 (r \ep_f ) +  \label{eq:photonWF}\\
  && \qquad \qquad \qquad \qquad + \de_{h+} \de_{\bar{h}+} m_f K_0 (r \ep_f) \Big) \nonumber
\end{eqnarray}
with
\begin{equation}
\ep_f = \sqrt{z(1-z)Q^2 + m_f^2} .
\end{equation}
Here $\lambda = 0$ and $+$ denote the longitudinal and transverse
wavefunctions respectively, $f$ denotes the quark flavour, and $K_0$
and $K_1$ are modified Bessel functions. $e_f$ is the electric charge
of the quark in units of the proton charge and $m_f$ the effective
mass of the quark.

\subsubsection{Smaller $Q^2$}

For smaller $Q^2$ the photon has also a hadronic component.  In
\cite{Avsar:2007ht} it was shown that also the total $\gamma^\star p$
cross section at HERA could be well described over a wide range of
energies and virtualities, when the hadronic component was simulated
by a relatively small effective quark mass $\approx 60$ MeV.  For the
exclusive reactions studied here we need a more careful treatment of
the hadronic component, and we expect that these processes can provide
relevant constraints on the photon wavefunction.  The hadronic
component should be particularly important for the real photons
produced in Deeply virtual Compton scattering (DVCS).

For small $Q^2$ a small effective quark mass allows for rather large dipoles 
with a corresponding large cross section. In the present analysis we 
include an improved description of confinement effects in the dipole
evolution (see section \ref{sec:confinement}), and we will therefore 
try to include confinement effects also in the photon wavefunction.
Our photon model is inspired by the Vector Meson Dominance modeling
introduced by Forshaw \emph{et al.}\ in \cite{Forshaw:2003ki} (which, in
turn, was inspired by \cite{Frankfurt:1997zk}). This model contains an
enhancement factor for dipoles of a typical hadronic size, together
with a large quark mass which suppresses dipoles larger than the confinement
scale. In our model we will use the same enhancement factor, but we use a 
suppression of large dipoles related to the confinement scale \rmax\ , 
instead of the large quark mass used in ref. \cite{Forshaw:2003ki}.

The actual implementation in our MC program relies on shrinking
dipoles larger than \rmax\ by reducing
the size $r\sub{pert}$ generated according to the perturbative photon
wavefunction to $r_{\text{soft}}$, defined by
\begin{equation}
  r_{\text{soft}}(r_{\text{pert}}) =
  R_{\text{shrink}} \sqrt{ \ln \left( 1 + \frac{r^2_{\text{pert}}}{R^2_{\text{shrink}}} \right) }.\label{eq:shrink}
\end{equation}
For small dipoles this gives $r_{\text{soft}} \approx r_{\text{pert}}$,
but for large dipoles it gives a Gaussian suppression. The parameter
$R_{\text{shrink}}$ is adjusted to give the same effective cutoff, $\rmax$,
as the one obtained for large dipoles in the cascade evolution. This
is obtained for $R_{\text{shrink}}=4.3\, \mathrm{GeV}^{-1}$. 

The enhancement factor for dipoles with a typical hadronic size, 
introduced in ref.\cite{Forshaw:2003ki}, is given by the form
\begin{equation}
  f(r) = \frac{1 + B_V\exp(-(r-R_V)^2/w_V^2)}{1 + B_V\exp(-R_V^2/w_V^2)}
  \label{eq:resfnc}
\end{equation}
This factor multiplies the squared photon wavefunction after the shift in
\eqref{eq:shrink}. The enhancement resembles very much the shape we
use for the proton wavefunction in \eqref{eq:pwave}, and we can think
of the the whole correction
\begin{equation}
  \abs{\Ps_\ga(r_{\text{pert}})}^2 \rightarrow
  \abs{\Ps_\ga(r_{\text{soft}})}^2 f(r_{\text{soft}})
  \label{eq:inphmo}
\end{equation}
as the modeling of the virtual photon fluctuating into vector meson
states with $r$-values of a hadronic scale. Partly this enhancement
can be thought of as due to a longer lifetime of these states, and partly
a simulation of a gluonic component in the vector meson, in a way
similar to our model of the initial proton wavefunction in 
section \ref{sec:inprmo}. The photon model contains three adjustable
parameters, $B_V$, $R_V$, and $w_V$, which have to be determined from
experiments.

\subsection{Meson wavefunctions}
\label{sec:mesonWF}

The wavefunction of a vector meson cannot be calculated
perturbatively, and has to be described by models. In the rest frame
it is generally assumed that the lowest Fock state with a single
$q\bar{q}$ pair dominates. This component must then be normalized to
1, in contrast to the photon for which the $q\bar{q}$ state is a
perturbative fluctuation. In addition the wavefunction at the origin
is determined by the decay rate of the vector meson. Thus there are
two constraints allowing two parameters in an ansatz to be
determined. In a boosted frame higher Fock states may then be
generated by gluon emission. Different models can differ in the
functional form used for the wavefunction in the rest frame, and in
the description of the transition from coordinate space to the
momentum fractions $z$ and $1-z$ used in the light-cone
wavefunction. We will here concentrate on the DGKP model
\cite{Dosch:1996ss} and the boosted Gaussian model
\cite{Forshaw:2003ki}, which in the analysis by Forshaw \emph{et
  al.}~give the best agreement with the experimental data.

\subsubsection{The DGKP model}
In this model for the meson wavefunction, proposed by Dosch, Gousset,
Kulzinger, and Pirner \cite{Dosch:1996ss}, it is assumed that the
dependence on the transverse and longitudinal coordinates, $r$ and
$z$, factorizes.  The transverse part of the wavefunction is assumed
to be a pure Gaussian, consistent with soft hadron--hadron
collisions. For the longitudinal component it assumes the form
proposed by Wirbel, Stech and Bauer \cite{Wirbel:1985ji}.  The
resulting light-cone wavefunction has the following form:
\begin{eqnarray}
  \Ps^{V 0}_{f h \bar{h}}(r,z) &=& \mathscr{N}_0 M_V \de_{-h\bar{h}}z(1-z)\sqrt{z(1-z)} \exp \left( -\frac{r^2 \om_L^2}{2} \right) \exp \left( - \frac{M_V^2 (z-0.5)^2}{2\om_L^2} \right) \nonumber\\
  \Ps^{V +}_{f h \bar{h}}(r,z) &=& \mathscr{N}_+ \left( \om_T^2 r i e^{i \te} ( z \de_{h+}\de_{\bar{h}-} - (1-z) \de_{h-}\de_{\bar{h}+} ) + m_f  \right) \sqrt{z(1-z)} \times \label{eq:DGKPWF}\\
  &&\exp \left( -\frac{M_V^2 (z-0.5)^2}{2 \om_T^2} \right) \exp \left( -\frac{r^2\om_T^2}{2} \right) . \nonumber
\end{eqnarray}
Here $M_V$ is the mass of the vector meson, and the size parameter
$\om$ and the normalization constant $\mathscr{N}$ are determined from
the electronic decay rate and the normalization condition. (Our
notation differs from the original paper, as we have collected the
multiplicative factors in the normalization constant $\mathscr{N}$.)
The shape of the wavefunction of the $\rho$ with the parameters we have
used (see table~\ref{tab:wfpar}) can be seen in figure \ref{fig:DGKP}.

\FIGURE[t]{
  \includegraphics[angle=0, scale=0.7]{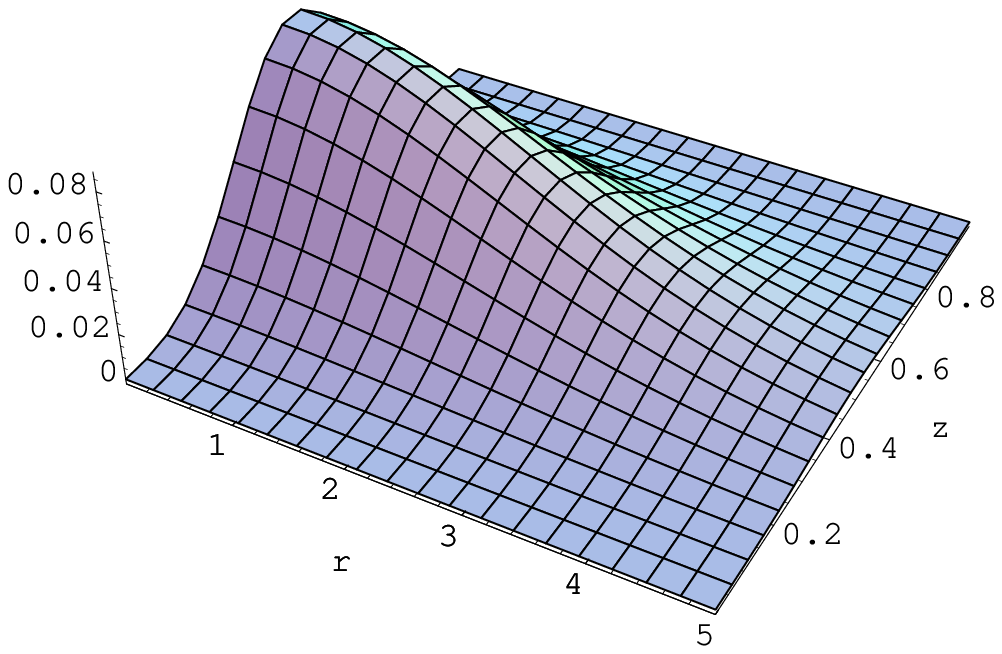}
  \includegraphics[angle=0, scale=0.7]{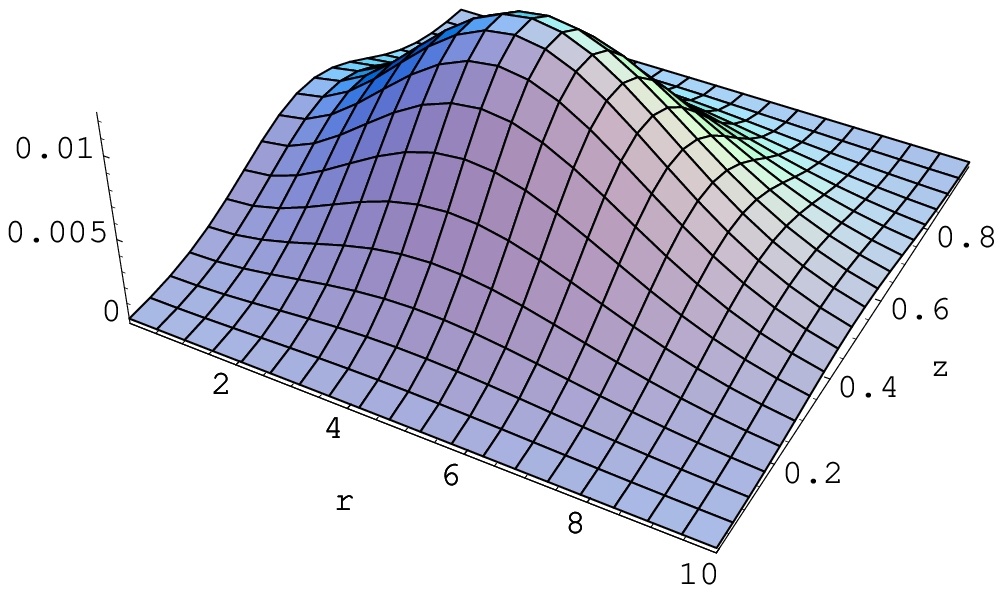}
  \caption {\label{fig:DGKP} The wavefunctions $\abs{\Ps_L(r,z)}^2$
    (left) and $\abs{\Ps_T(r,z)}^2$ (right) of the DGKP model for a
    $\rho$ meson with our quark mass of 60~MeV. Note the different scales
    in $r$, both in GeV$^{-1}$. } }

\TABLE[t]{
  Boosted Gaussian\hfil DGKP\\[2mm]
  \begin{tabular}{|c||c|c|c|c|}
    \hline
    $V$ & $M_V$ & $m_{uds}$ & $m_c$ & $R^2$ \\
    \hline
    $\rho$ & 0.77 & 0.06 & 1.4 & 12.3 \\
    $\phi$ & 1.02 & 0.06 & 1.4 & 2.44 \\
    $\psi$ & 3.1 & 0.06 & 1.4 & 10 \\
    \hline
  \end{tabular}\hfil
  \begin{tabular}{|c||c|c|c|c|c|}
    \hline
    $V$ & $M_V$ & $m_{uds}$ & $m_c$ & $\omega_L$ & $\omega_T$ \\
    \hline
    $\rho$ & 0.77 & 0.06 & 1.4 & 0.33 & 0.22 \\
    $\phi$ & 1.02 & 0.06 & 1.4 & 0.37 & 0.26 \\
    $\psi$ & 3.1 & 0.06 & 1.4 & 0.69 & 0.56 \\
    \hline
  \end{tabular}
  \caption{\label{tab:wfpar} The parameters used for the boosted
    Gaussian and DGKP wavefunctions in this paper in GeV-based units.}}

\subsubsection{The boosted Gaussian model}
The ``boosted'' models are obtained by assuming a given wavefunction
in the meson rest frame. This is then boosted into a light-cone wavefunction using the Brodsky-Huang-Lepage prescription
\cite{Brodsky:1980vj}, in which the invariant mass of the
quark-antiquark pair is the same in the rest frame and the light-cone
frames. The result of this procedure is not factorizing in $r$ and
$z$. In the simplest version the initial wavefunction in the rest
frame is a simple Gaussian. In an alternative version by Nemchik
\emph{et al.}\ \cite{Nemchik:1996cw} a hard Coulomb contribution is added,
dominating for small $r$. For the pure Gaussian version suggested by Forshaw \emph{et al.}, which we
assume in this analysis, the resulting wavefunction has the following
form
\begin{eqnarray}
  \Ps^{V 0}_{f h \bar{h}}(r,z) &=& \frac{\mathscr{N}_0}{M_V}
  \left( z(1-z)M_V^2 + m_f^2 + 8\frac{z(1-z)}{R^2} - 
    \left( 4 \frac{z(1-z)r}{R^2} \right)^2 \right) \de_{h \bar{h}}
  \times \nonumber\\
  &&\exp \left( -\frac{m_f^2 R^2}{8z(1-z)} \right)
  \exp \left( -2z(1-z)\frac{r^2}{R^2} \right) 
  \exp \left( \frac{R^2}{2} m_f^2 \right)  \label{eq:bgWF}\\
  \Ps^{V +}_{f h \bar{h}}(r,z) &=& \mathscr{N}_+
  \left( 4z(1-z)\frac{r}{R^2} ie^{i\te}
    ( z \de_{h+} \de_{\bar{h}-} -
    (1-z) \de_{h-} \de_{\bar{h}+} ) + m_f \de_{h+} \de_{\bar{h}+} \right)
  \times \nonumber\\
  && \exp \left( -\frac{m_f^2 R^2}{8z(1-z)} \right)
  \exp \left( -2z(1-z)\frac{r^2}{R^2} \right)
  \exp \left( \frac{R^2}{2} m_f^2 \right) \nonumber
\end{eqnarray}
In this model the transverse size $R$ of the meson, and the
normalization $\mathscr{N}_\la$ are the two parameters to be
tuned. (In our notation all multiplicative constants have also here
been included in a single normalization constant.) The shape of the
wavefunction for the $\rho$ meson using the parameters in
table~\ref{tab:wfpar} is shown in figure \ref{fig:Boosted}.

\FIGURE[t]{
  \includegraphics[angle=0, scale=0.7]{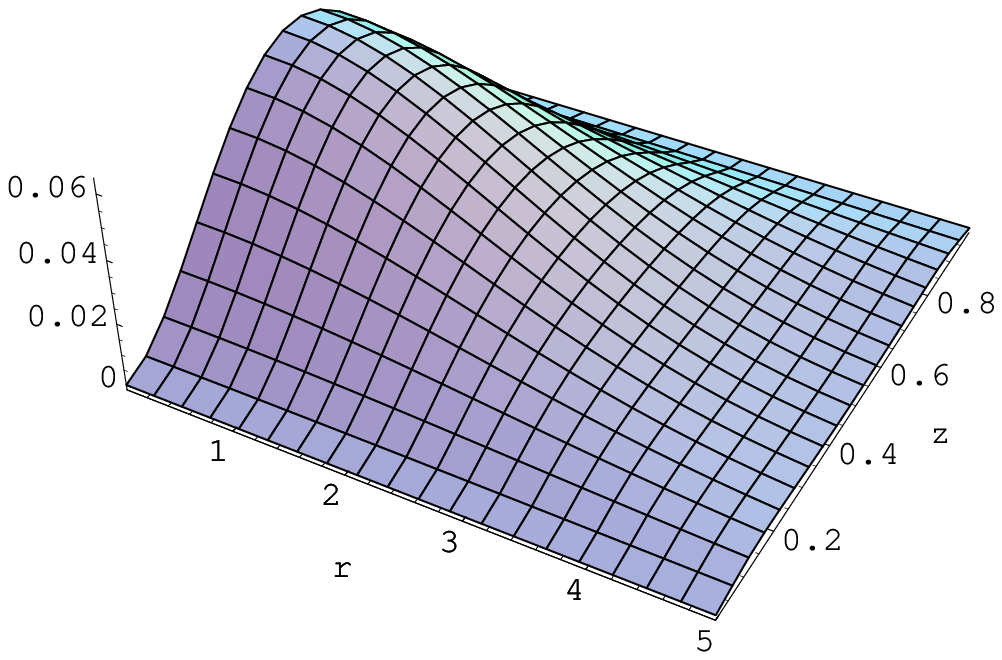}
  \includegraphics[angle=0, scale=0.7]{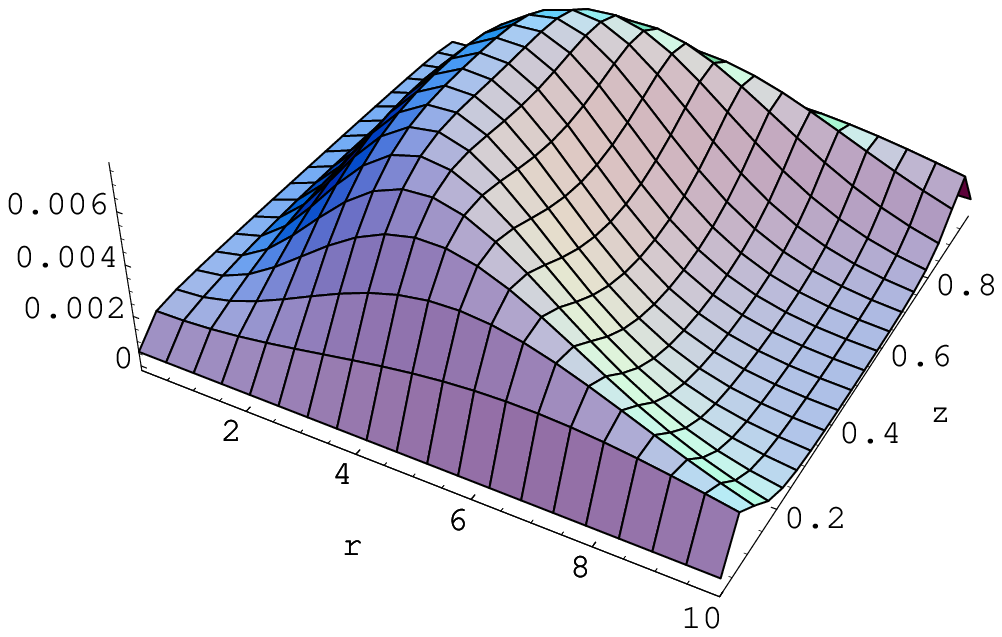}
  \caption {\label{fig:Boosted} The wavefunctions $\abs{\Ps_L(r,z)}^2$
    (left) and $\abs{\Ps_T(r,z)}^2$ (right) of the Boosted Gaussian
    model for a $\rho$ meson with our quark mass of 60~MeV. Note the
    different scales in $r$, both in GeV$^{-1}$. } }

\section{Tuning of  parameters and the differential \boldmath$pp$ cross section}
\label{sec:tuning}

\subsection{The total and elastic  \boldmath$pp$ cross section}
\label{sec:total-elastic-pp}

We start by tuning the model to $pp$ scattering data.
Here the model contains 4 parameters, $\lam$ and $r_{\max}$ describing
the dipole evolution, and $R_p$ and $w$ determining the proton wave function
in \eqref{eq:pwave} (with $C$ fixed by normalization).
In  ref.~\cite{Avsar:2007xg} we found that the values for $\lam$ and $r_{\max}$
are correlated, such that a larger $r_{\max}$ can be compensated by a smaller 
$\lam$. It was also noted that the integrated elastic cross section favors a
narrow proton wave function, corresponding to a small value for the parameter
$w$. A large $w$-value, or a single Gaussian $\propto
\exp(-\mathbf{r}^2/R_p^2)$, gives too large fluctuations and correspondingly a
too small elastic cross section. 
To constrain the fit we here add 
the differential cross section $d\sigma/dt$ to the integrated
cross sections $\sigma\sub{tot}$, $\sigma\sub{diff}$, and $\sigma\sub{el}$
studied in \cite{Avsar:2007xg}. We will then in next section check if 
the result also
can reproduce the quasi-elastic cross sections in $\gamma^*p$ collisions.

With the proton wavefunction given by \eqref{eq:pwave} the total and elastic cross sections are given by
\begin{equation}
  \si\sub{tot} = 2 \int d^2 \bl{b} d^2 \bl{r}_{p1} d^2 \bl{r}_{p2}
  \abs{\Psi_p(\bl{r}_{p1})}^2 \abs{\Psi_p(\bl{r}_{p2})}^2 \langle 1-e^{-F}\rangle_{12}, 
  \label{eq:pptotCS}
\end{equation}
\begin{equation}
  \si\sub{el} = \int d^2 \bl{b}
  \abs{\int  d^2 \bl{r}_{p1} d^2 \bl{r}_{p2} \abs{\Psi_p(\bl{r}_{p1})}^2 \abs{\Psi_p(\bl{r}_{p2})}^2
    \langle 1-e^{-F}\rangle_{12}}^2.\label{eq:ppelCS}
\end{equation}
Here $\bl{b}$ is the impact parameter, $\bl{r}_{pi}$ ($i=1,2$) parameterizes
the size and orientation of the triangles describing the colliding protons.
The Monte Carlo is used to simulate the dipole evolution in the
rest frame of the collision, and to calculate $1-e^{-F}$. 
The average $\langle 1-e^{-F}\rangle_{12}$ is over simulations for 
fixed impact parameter and starting dipole states $\bl{r}_{1}$ and $\bl{r}_{2}$.
Note that in the elastic
cross section the average over evolutions and the integrals over
the wave functions is taken on amplitude level before taking the absolute square.

When tuning the parameters we find that all observables are almost 
independent of $w$ below 0.5~GeV$^{-1}$. We therefore
decided to neglect the fluctuations in the proton
size completely and set the width to zero, turning the proton
wavefunction into a delta function at $R_p$.

\FIGURE[t]{
  \begin{minipage}{1.0\linewidth}
    \begin{center}
      \includegraphics[angle=0, scale=1.5]{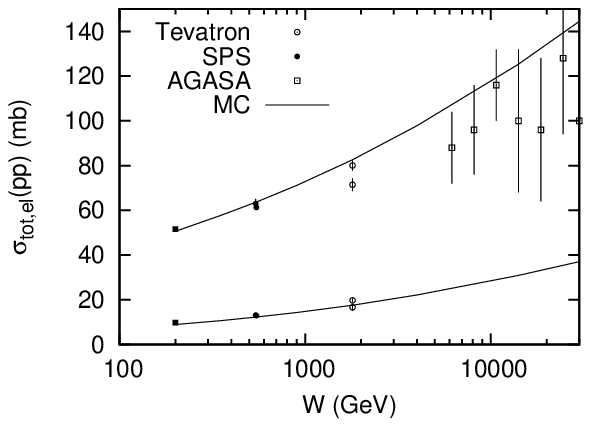}
    \end{center}
  \end{minipage}
  \caption {\label{fig:ppEdep} The total and elastic cross section for
    $pp$ collision. The upper cross sections are total cross sections,
    while the lower cross sections are the elastic ones. Tevatron data
    are from \cite{Abe:1993xx,Abe:1993xy,Amos:1990jh,Avila:2002bp}, SPS
    data are from \cite{Augier:1994jn} and cosmic ray data are from
    \cite{Block:2000pg}. The lines are our model with tuned parameters. } }

If the total and integrated elastic cross sections are tuned at one 
energy, we find that the energy dependence of these cross sections depends
very weakly, if at all, on the parameters of the model. Thus this energy
dependence cannot be tuned, and the fact that 
it is close to the experimental results 
is therefore a direct consequence of the model. 
Our results for the
total and elastic $pp$ cross sections can be seen in figure
\ref{fig:ppEdep}.

Extrapolating to higher energies we find the total cross
section at the LHC nominal energy, 14~TeV, to be about 125~mb (117~mb at
10~TeV). We note that this is a rather high value compared to other
predictions. Thus the Donnachie-Landshoff parameterization gives
101.5~mb at 14~TeV \cite{Donnachie:1992ny}, while an analysis by Khoze, 
Martin, and Ryskin gives about 90~mb \cite{Ryskin:2007qx}.
The predicted elastic cross section is about
31~mb for the LHC at 14~TeV (28~mb at 10~TeV).

\FIGURE[t]{
  \includegraphics[angle=0, scale=2.5]{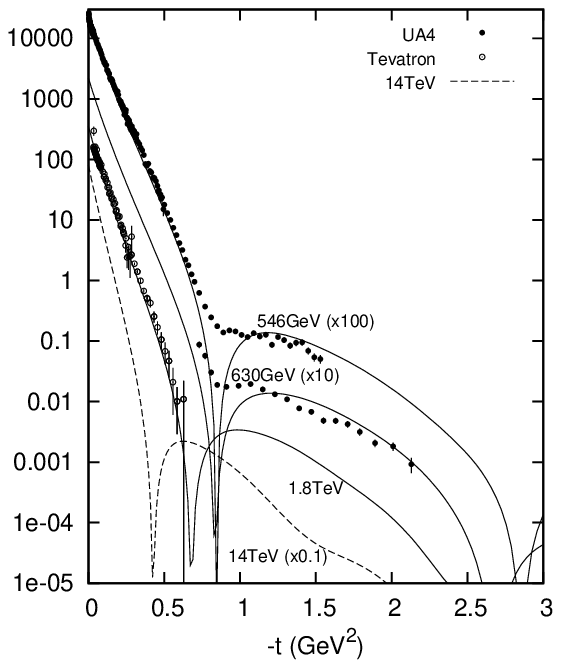}
  \caption {\label{fig:pptdep} The elastic cross section as function
    of $t$ in mb/GeV$^2$.  The numbers in parenthesis indicate how the
    data has been scaled. The lines are our model with tuned
    parameters. Predictions for 14~TeV is also included. Data are from
    \cite{Abe:1993xx,Amos:1990fw}, \cite{Bernard:1986ye} and
    \cite{Adamus:1987sx}. } }

\subsection{The differential elastic \boldmath$pp$ cross section}

The differential elastic $pp$ cross section is given by
\begin{equation}
  \frac{\si\sub{el}}{dt} = \frac{1}{4\pi} \abs{\int d^2 \bl{b} e^{-i \bl{q}
      \cdot \bl{b}}  d^2 \bl{r}_{p1} d^2 \bl{r}_{p2}
    \abs{\Psi_p(\bl{r}_{p1})}^2 \abs{\Psi_p(\bl{r}_{p2})}^2 \langle 1-e^{-F}\rangle_{12}}^2,
    \,\,\,\mathrm{with}\,\,\,t = -\bl{q}^2.
  \label{eq:ppdiffCS}
\end{equation}
We here neglect the real part of the amplitude, and therefore $d\si
/dt$ will have zeroes from the Fourier transform of the amplitude in
\eqref{eq:ppdiffCS}. Even though the true complex amplitude will
not be identically zero, the real part is still assumed to be small,
producing a dip at some value $t=t_0$, related to the inverse square
of the size of the proton at the relevant energy. This dip is visible
in some of the experimental data shown in figure \ref{fig:pptdep},
where we have also included the results from the simulations. The
parameters which are most sensitive to these distributions are $R_p$,
which determines the size of the proton at rest, and $\rmax$ which
regulates the maximal size of new dipoles, and therefore the increase with
energy of the proton radius and
the variation of the dip position. However, the
slope of the distribution is basically independent of our parameters,
as is the cross section at large $t$-values. Nevertheless, we are able
to get a very good description of the data at all $t$-values even
though the cross sections vary over many orders of magnitude. In
figure \ref{fig:pptdep} we also show our result for the LHC, which
predicts the location of the dip in the $t$-dependence at 0.43~GeV$^2$
at $\sqrt{s}=14$~TeV (0.47~GeV$^2$ at 10~TeV).  The values of the tuned
parameters can be found in table \ref{tab:tunedpar}.

\TABLE[t]{
  \begin{minipage}{1.0\linewidth}
    \centering
    \begin{tabular}{|c|r|}
      \hline
      \lam & $0.2$~GeV\\
      $r_{\max}$ & $2.9$~GeV$^{-1}$\\
      $R_p$ & $3.0$~GeV$^{-1}$\\
      $w$ & $0$~GeV$^{-1}$\\
      \hline
    \end{tabular}
  \end{minipage}
  \caption{The tuned values the parameters for the proton wavefunction
    and the perturbative evolution used for our model throughout this paper.}
  \label{tab:tunedpar}}

\subsection{The total \boldmath$\gamma^\star p$ cross section and tuning
the photon wave function}
\label{sec:total-gamma-p}

We will here use $\Ps_\ga(Q,\bl{r},z)$ to denote the photon
wavefunction in \eqref{eq:inphmo}, where for small $Q^2$ the perturbative 
wavefunction is modified to account for the hadronic component of the photon.
The total $\gamma^*p$ cross section can be written

\begin{equation}
  \si_{tot}(\gamma^*p) = \sum_{\la f} \int d^2 \bl{b} d^2 \bl{r}_p d^2 \bl{r} 
  dz \abs{\Psi_p(\bl{r}_p)}^2 \abs{\Ps^{\la}_{\ga , f}(Q,\bl{r},z)}^2
  \langle 1-e^{-F}\rangle_{dp},
\end{equation}
where $\la$ is the polarization of the photon and $f$ is the flavour
of the quark-antiquark pair created by the photon. $\langle
1-e^{-F}\rangle_{dp}$ is now an average of the evolution of the dipole
from the photon side and of the dipoles from the proton side. It
depends not only on the total energy, the size of the proton and
photon dipoles and $b$, but has also a weak dependence on $z$.

The three parameters $B_V$, $R_V$ and $w_V$ in the enhancement
factor in \eqref{eq:resfnc} were fitted to the total
$\gamma^\star p$ cross section as measured at HERA.
Here the value of $R_V$ determines the range in $Q^2$ where the enhancement is
significant, while $w_V$ determines how fast it falls off for 
large $Q^2$. The parameter $B_V$ is
just an overall strength of the hadronic component.

A good fit to data was obtained with a wave function for the hadronic
component similar to the proton wave function, having a size 
$R_V\approx$ 3~GeV$^{-1}$ and a small width.
The total $\gamma^\star p$ cross section with and without the effects
of confinement and vector meson contributions are shown in figure
\ref{fig:gptot}. The tuned values
are given in table~\ref{tab:tunedVMDpar}. These parameters are quite
different from the ones used by Forshaw \emph{et al.}, who had an $R_V$ of
6.84~GeV$^{-1}$ and a fairly large width \cite{Forshaw:1999uf},
which thus gives a stronger enhancement for large dipoles. For large dipole
sizes the elementary dipole-proton cross section assumed in 
ref. \cite{Forshaw:2003ki} are also significantly different from
the corresponding ones in our model. The reason why we anyhow can get similar
results is that in \cite{Forshaw:2003ki} the very large dipoles are
suppressed by a large quark mass, and the enhancement therefore not very 
effective. A reason for a smaller width in our wavefunction is also that
fluctuations in the dipole cascade are included in our formalism,
which is compensated by less fluctuations in the wavefunction.

We also note that the cross sections presented in fig. \ref{fig:gptot} are
somewhat smaller than the corresponding results in refs. 
\cite{Avsar:2007xg} and \cite{Avsar:2007ht}. This is a consequence of the
more consistent treatment of confinement in the present analysis,
which gives a stronger suppression for larger dipoles. We believe, however,
that a much stronger test of the hadronic component will rely on the
results for the quasi-elastic reactions discussed in next section.

\FIGURE[t]{
  \includegraphics[angle=0, width=0.5\linewidth]{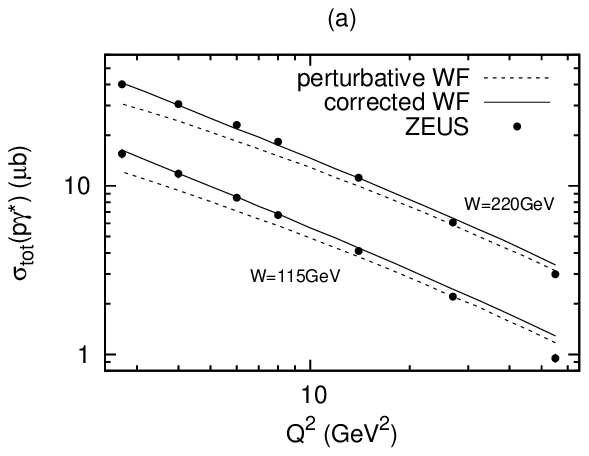}%
  \includegraphics[angle=0, width=0.5\linewidth]{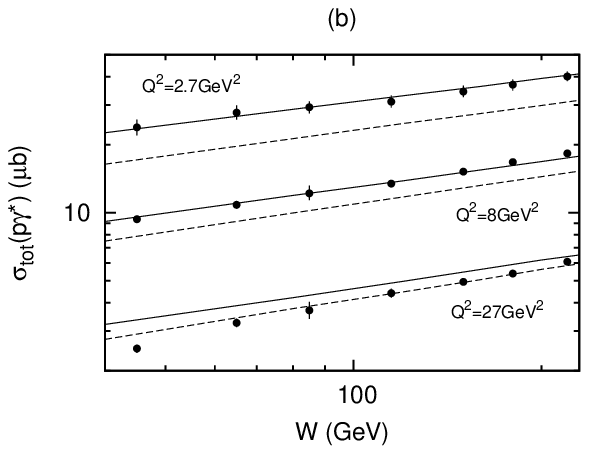}
  \caption {\label{fig:gptot} The total cross section of photon--proton
    collision as function of photon virtuality and center-of-mass
    energy. The dashed line is
    calculated with a purely perturbative photon wavefunction, while
    the full line is with a photon wavefunction with both confinement
    and VMD corrections. Experimental data are from
    \cite{Chekanov:2005vv} } }

\TABLE[t]{
  \begin{minipage}{1.0\linewidth}
    \centering
    \begin{tabular}{|c|l|}
      \hline
      $R_V$ & $3.0$~GeV$^{-1}$\\
      $w_V$ & $0.2$~GeV$^{-1}$\\
      $B_V$ & $9.0$\\
      \hline
    \end{tabular}
  \end{minipage}
  \caption{The tuned values of the parameters of the vector meson
    resonance function $f(r)$ used for our model throughout this
    paper.}
  \label{tab:tunedVMDpar}}

\section{Results for quasi-elastic \boldmath$\gamma^\star p$ collisions}
\label{sec:results-quasi-elast}

In this section we will study predictions for quasi-elastic $\gamma^* p$
collisions, using the photon wavefunction parameters determined in 
section \ref{sec:total-gamma-p}.

\subsection{Deeply Virtual Compton Scattering}
\label{sec:dvcs}

In Deeply Virtual Compton Scattering, DVCS, the incoming particle
is a virtual photon, while the outgoing particle is a
real photon with wavefunction $\Ps^{\la}_{\ga , f}(0,\bl{r},z)$. 
According to eqs. (\ref{eq:amplitude}, \ref{eq:elastic}) the
cross section is given by
\begin{equation}
  \si\sub{DVCS} = \int d^2 \bl{b} \sum_{\la}
  \abs{ \int  d^2 \bl{r}_p d^2 \bl{r} dz \sum_{f}  \abs{\Psi_p(\bl{r}_p)}^2
    \Ps^{\star\ga\la}_{f}(Q,\bl{r},z) \Ps^{\ga\la}_{f}(0,\bl{r},z)
    \langle 1-e^{-F}\rangle_{dp}}^2.
\end{equation}
The differential cross section $d\si/dt$
is obtained from the Fourier transform as shown in \eqref{eq:diffelastic}.
The results obtained using the parameter values in table \ref{tab:tunedVMDpar}
are presented in figures \ref{fig:dvcs} and \ref{fig:compton}.
We see that the results from the model indeed agrees with the data, both in
normalization and in the dependence on  $Q^2$, $W$ and $t$.
As this quasi-elastic reaction is very sensitive to the fluctuations and the
transverse size of the hadronic component of the real photon, this is a
clear support for the proton-like wave function. 

\FIGURE[t]{
  \includegraphics[angle=0, width=0.5\linewidth]{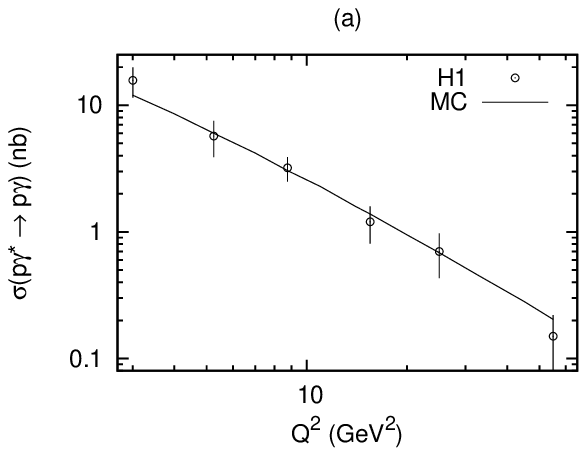}%
  \includegraphics[angle=0, width=0.5\linewidth]{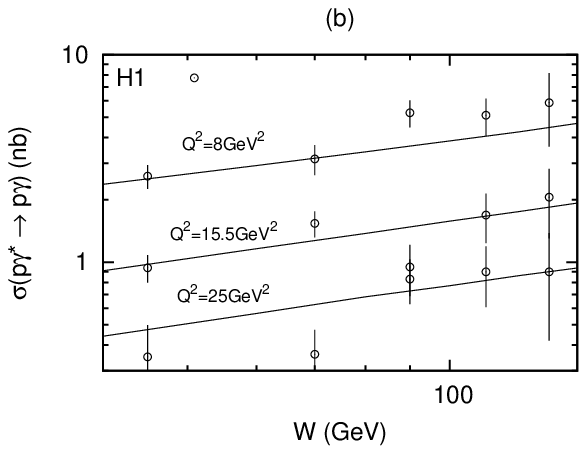}
  \caption {\label{fig:dvcs} The cross section of $\gamma^\star p\to
    \gamma p$ for $W = 82$~GeV as function of $Q^2$ (a) and as function of $W$ for three different $Q^2$ (b). Experimental data are from
    \cite{Aktas:2005ty,:2007cz}. } }

\FIGURE[t]{
  \begin{minipage}{1.0\linewidth}
    \begin{center}
  \includegraphics[angle=0, width=0.5\linewidth]{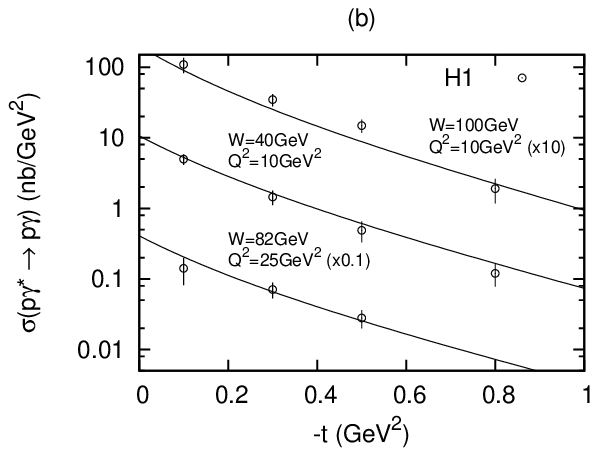}
    \end{center}
  \end{minipage}
  \caption {\label{fig:compton} The cross section of $\gamma^\star p\to \gamma p$ as function of $t$. The brackets refers to
    $(W,Q^2)$. Note that the three series have been scaled by
    10,1 and 0.1 for better readability. Experimental data are from
    \cite{:2007cz}. } }

\subsection{Exclusive Production of Light vector Mesons}
\label{sec:rho}
 
The cross section for exclusive vector meson production, 
$\gamma^\star p\to Vp$, can be calculated
in exactly the same way as for DVCS, only replacing the real photon 
wavefunction by a meson wavefunction:
\begin{equation}
  \si\sub{el} = \int d^2 \bl{b} \sum_\la
  \abs{ \int d^2 \bl{r} d^2 \bl{r}_p  dz \sum_{f}  \abs{\Psi_p(\bl{r}_p)}^2
    \Ps^{\star\ga\la}_{f}(Q^2,\bl{r},z) \Ps^{V\la}_{f}(\bl{r},z)
    \langle 1-e^{-F}\rangle_{dp}}^2 .
  \label{eq:VMCS}
\end{equation}

As before we have ignored the real part of the amplitude. Contrary to
the case of $pp$ scattering it has been shown in \cite{Forshaw:2003ki} that in
exclusive production of light vector mesons the real part can be quite
large, for large $Q^2$ or large $W$ as large as one half of 
the imaginary part. This would mean that we underestimate the cross 
section by up to
25\% in these regions. However, compared to the uncertainties in the
vector meson wavefunctions, this error is small.

From our tuning of the hadronic part of the photon wavefunction, it could
seem natural to assume some universal hadron size and maybe try to
model the vector meson wavefunctions as a simple gluon--gluon dipole
with a size of $3$~GeV$^{-1}$ and a small width. However, this would
not naturally give us a $z$-dependence and we would not include the
possibility that the
vector meson may fluctuate into a photon, which could correspond to an
enhancement at small $r$-values. Therefore we will simply use the
boosted Gaussian and DGKP wavefunctions introduced in section
\ref{sec:mesonWF} to estimate the $\gamma^\star p\to Vp$
cross section. Throughout we will use the parameters
listed in table~\ref{tab:wfpar}.

As before, the $t$-dependence of the cross section can be calculated
through a Fourier transform of the amplitude. We are also able to
calculate the ratio between the longitudinal to the transverse
cross sections and compare with experimental data.

\FIGURE[t]{
  \includegraphics[width=0.5\linewidth]{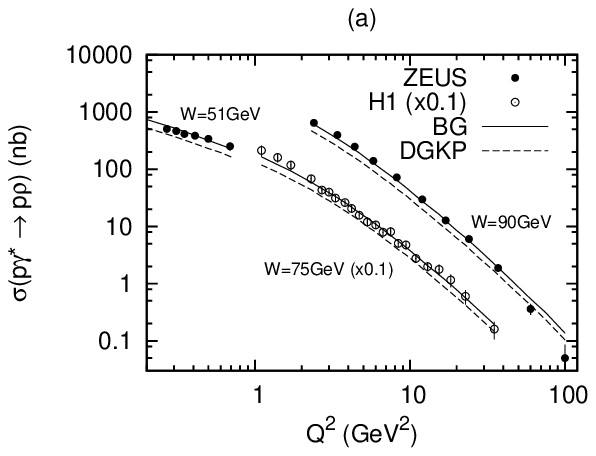}%
  \includegraphics[width=0.5\linewidth]{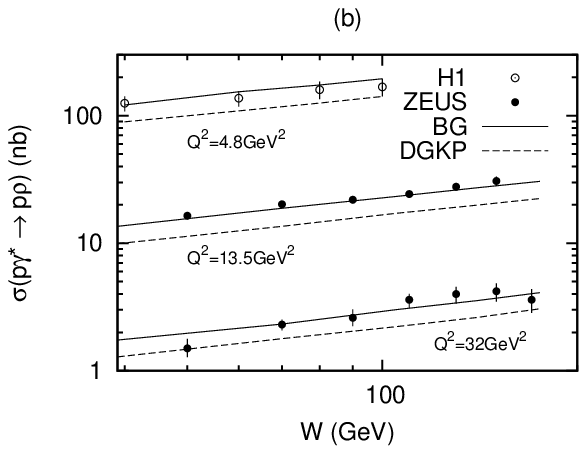}
  \caption{\label{fig:rhoQdep} The cross section for $\gamma^\star
    p\to \rh p$. (a) As function of the photon virtuality. The H1 data has been
    moved down a factor 10 for better readability. The full and dashed
    line are with the two different meson wavefunction described in
    \ref{sec:mesonWF}. (b) As function of the center-of-mass energy $W$.
    Experimental data are from
    \cite{Adloff:1999kg,Chekanov:2007zr,Breitweg:1998nh}. } }

\FIGURE[t]{
  \includegraphics[width=0.5\linewidth]{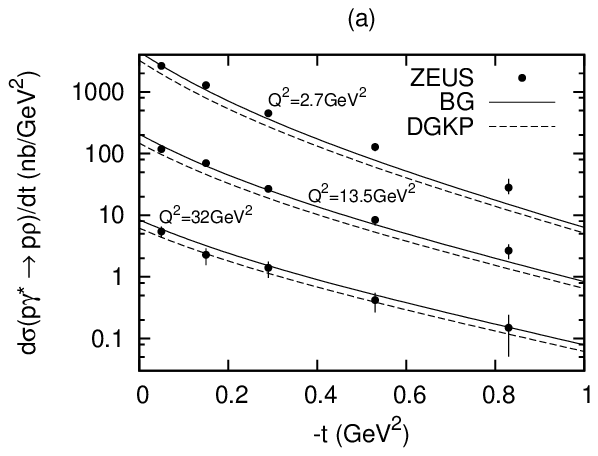}%
  \includegraphics[width=0.5\linewidth]{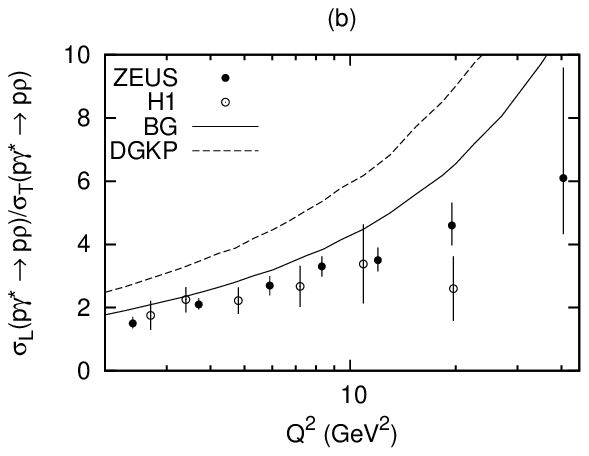}
  \caption{\label{fig:rhotdep} (a) The differential cross section for
    $\gamma^\star p\to \rh p$ as function of transferred momentum
    squared $\abs{t}$. Three different $Q^2$ has been used, all with a
    center-of-mass energy of 90~GeV. (b) The ratio of longitudinal and
    transverse cross section for $\gamma^\star p\to \rh p$ as function
    of the photon virtuality. Experimental data are from
    \cite{Chekanov:2007zr,Adloff:1999kg}.} }

Starting with $\rho$ meson production, the results are shown in figures
\ref{fig:rhoQdep} and \ref{fig:rhotdep}. We see that the model calculations
reproduce experimental
data rather well, including the dependence on virtuality $Q^2$, energy $W$ and
momentum transfer $t$. The Boosted Gaussian
wavefunction is in general providing the closer fit, while DGKP is having
problems mainly in the ratio between longitudinal and transverse cross
sections. It should be noted, however, that this ratio could be changed
if we decided to use different parameters for the resonance function
in \eqref{eq:resfnc} for the different polarizations.

\FIGURE[t]{
  \includegraphics[width=0.5\linewidth]{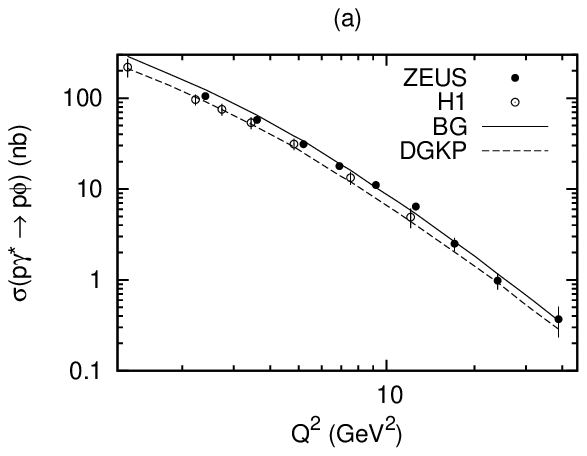}%
  \includegraphics[width=0.5\linewidth]{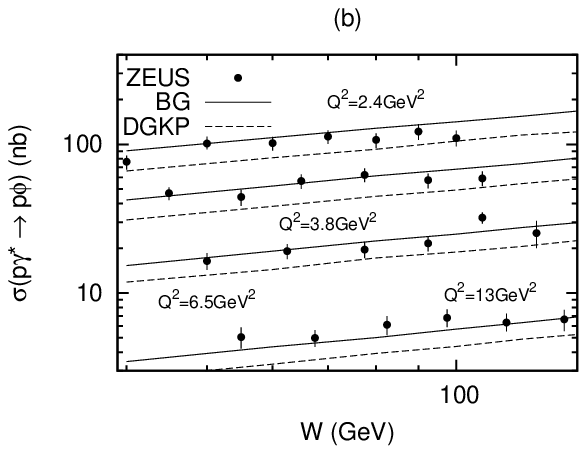}
  \caption {\label{fig:phiQdep} The cross section for $\gamma^\star
    p\to \ph p$ (a) As function of the photon virtuality at
    center-of-mass energy 75~GeV. The full and dashed line are with
    the two different meson wavefunction described in section
    \ref{sec:mesonWF}. (b) As function of the center-of-mass energy
    for four different photon virtualities $Q^2$. Experimental data
    are from \cite{Adloff:2000nx,Chekanov:2005cqa}. } }

\FIGURE[t]{
  \begin{minipage}{1.0\linewidth}
    \begin{center}
      \includegraphics[width=0.5\linewidth]{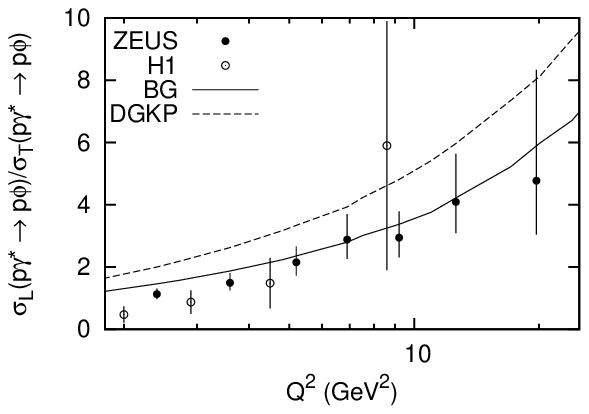}
    \end{center}
  \end{minipage}
  \caption {\label{fig:phiLoverT} The ratio of longitudinal and
    transverse cross section for $\gamma^\star p\to \ph p$ as function
    of the photon virtuality at a center-of-mass energy of
    75~GeV. Experimental data are from
    \cite{Adloff:2000nx,Chekanov:2005cqa}. } }

Also in $\phi$ production our model agrees well with experimental data, as can
be seen in figures \ref{fig:phiQdep} and \ref{fig:phiLoverT}. The
qualitative properties are similar to those of $\rho$ production

We note that the more stringent test of the hadronic component of the
photon is obtained from DVCS. The ratio between vector meson production and
DVCS is then more a test of the vector meson wavefunctions. It is therefore not
surprising that we here get results similar to those in
ref.~\cite{Forshaw:2003ki}. The $t$-dependence presented in 
fig.~\ref{fig:rhotdep} is, however, a result which in our model is 
sensitive to both the photon and vector meson wavefunctions, while
in ref.~ \cite{Forshaw:2003ki} it was fixed by experimental data.
From fig.~\ref{fig:rhotdep} we see that for lower $Q^2$ the slope in
the model is somewhat too steep, thus indicating a too wide wavefunction
for the $\rho$ meson. We see from figs. \ref{fig:DGKP} and \ref{fig:Boosted}
that the $\rho$ wavefunctions for
transverse polarization are extending well beyond 5 $\mathrm{GeV}^{-1}$. A
faster falloff for large $r$-values would here give a better agreement
with the observed $t$-dependence.


\subsection{\boldmath Exclusive $\psi$ Production}
\label{sec:jpsi}

In the case of $\psi$ production we necessarily encounter more
uncertainties. The result is sensitive to the mass of the charm quark,
and here we use the value 1.4~GeV, which in the analysis in
ref.~\cite{Avsar:2007ht} gave the correct charm contribution, $F_2^c$, to 
the proton structure
function. We have not included a $\psi$ component in the photon wave
function, although this would in principle be possible. For the $\psi$ meson
wavefunction we use the parameters in table~\ref{tab:wfpar}.

\FIGURE[t]{
  \includegraphics[width=0.5\linewidth]{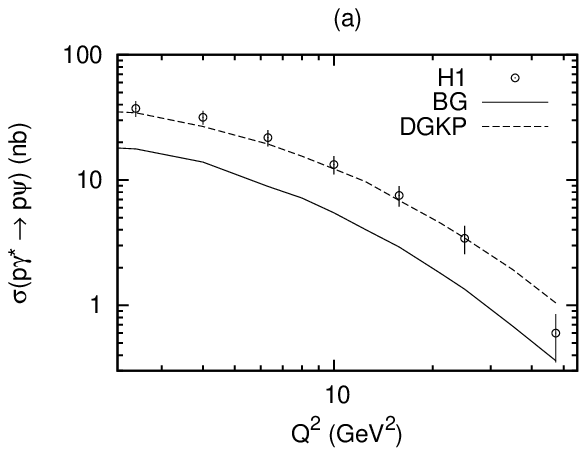}%
  \includegraphics[width=0.5\linewidth]{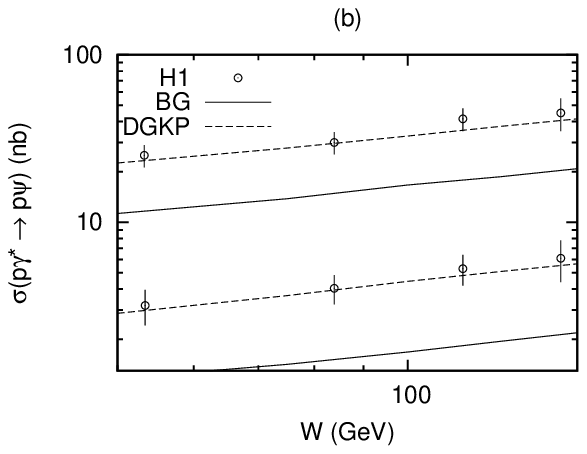}
  \caption {\label{fig:JpsiQdep} The cross section for $\gamma^\star
    p\to \psi p$ (a) As function of the photon virtuality at
    center-of-mass energy 90~GeV. The full and dashed line are with
    the two different meson wavefunction described in
    \ref{sec:mesonWF}. (b) As function of center-of-mass energy for
    $Q^2 =$ 22.4~GeV$^2$ (lower data) and 3.2~GeV$^2$ (upper
    data). Experimental data are from \cite{Aktas:2005xu}. } }

\FIGURE[t]{
  \begin{minipage}{1.0\linewidth}
    \begin{center}
      \includegraphics[width=0.5\linewidth]{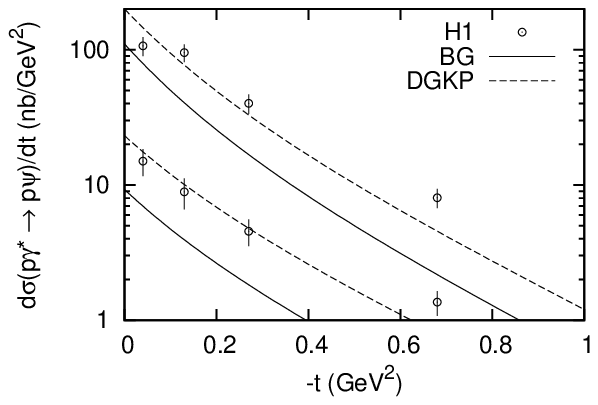}
    \end{center}
  \end{minipage}
  \caption {\label{fig:Jpsitdep} The cross section for $\gamma^\star
    p\to \psi p$ as function of transferred momentum squared $t$ for
    $Q^2 =$ 22.4~GeV$^2$ (lower data) and 3.2~GeV$^2$ (upper
    data). Experimental data are from \cite{Aktas:2005xu}. } }

Our results are presented in figures \ref{fig:JpsiQdep} (dependence on $Q^2$
and $W$) and \ref{fig:Jpsitdep} (dependence on $t$). We note that 
the Boosted Gaussian
wavefunction gives a too low cross section over the entire kinematic
region, while DGKP model agrees very well both in its normalization and
in its dependence on $Q^2$ and $W$. Both models show, however, a somewhat 
steeper $t$-dependence than the experimental data. This can be compared to the
corresponding result for $\rho$ meson production, and we conclude that 
also for the $\psi$ meson the parameters in table~\ref{tab:wfpar} gives
somewhat too wide wavefunctions.

\section{Conclusions and Outlook}
\label{sec:conc}

In this paper we have spent some effort on the modeling of
non-perturbative aspects of the proton, photon and vector meson
wavefunctions. None of our models are in any way unique or on solid
theoretical grounds. However, they do allow us to compare our dipole
evolution model directly to experimental data. Fixing the wavefunction
parameters at one energy we find 
that the energy dependence of both total and (quasi-)elastic
cross sections are well described by the cascade evolution,
and rather independent of our
modeling of the wavefunctions. Also the slope in $d\sigma/d t$ for
elastic $pp$ scattering and DVCS is in agreement
with experimental data independently of the tuning. This
indicates a very high predictive power of our evolution model both
when it comes to the average multiplicity and sizes of the dipoles
(mainly influencing the total cross sections), the rate of increasing
transverse size due to the dipole cascade (determining the energy variation of
the dip position), and the fluctuations
(mainly influencing the magnitude of elastic cross sections and their
$t$-dependence).

Nevertheless, our modeling of the non-perturbative wavefunctions does
give us valuable insights. Including the fluctuations in the cascade,
the fluctuations in the proton wavefunction have to be rather small,
in order to give the observed elastic cross section.
The photon wavefunction clearly needs a hadronic component with a wavefunction
with similar size as the proton and with similarly small fluctuations.
The fact that the size comes out to be the same as the
size of our proton may be a coincidence, but it could also indicate
that there is a universal size of hadrons, at least when consisting of
light quark flavours. 

For the vector meson wavefunctions we have tested two different forms,
which in the analysis by Forshaw \emph{et al.}\ gave the best agreement with
data for diffractive vector meson production. For light mesons the
best result was obtained by the boosted Gaussian wavefunctions, while for 
$\psi$ production the DGKP wavefunction was favored. In both cases
the $t$-dependence was somewhat too steep, indicating that these
wavefunctions extend out to too large $r$-values, where in particular
the wavefunctions for transverse polarization are much wider than our
wavefunctions for the proton and the photon.

The robustness of our model for dipole evolution, both for inclusive
and exclusive cross sections increases our confidence that it can be
used to also model fully exclusive final states. In future
publications we will therefore concentrate on turning our Monte Carlo
simulation program into a full-fledged event generator, which would
then be able to model multi-particle production at e.g.\ the LHC, with
special emphasis on the underlying event and multiple scatterings.

\section*{Acknowledgment}

We want to thank Emil Avsar for valuable discussions. This work is
supported in part by the Marie Curie research training network
``MCnet'' (contract number MRTN-CT-2006-035606) and by the Swedish
Foundation for International Cooperation and Higher Education --
STINT, contract number 2006/080. G\"osta Gustafson also acknowledges
support from the Deutsche Forschungsgemeinschaft.

\bibliographystyle{utcaps}
\bibliography{/home/shakespeare/people/leif/personal/lib/tex/bib/references,refs}

\end{document}